# Statistical-Raman-Microscopy and Atomic-Force-Microscopy on Heterogeneous Graphene Obtained after Reduction of Graphene Oxide

Siegfried Eigler,[*,1] Ferdinand Hof,[1] Michael Enzelberger-Heim,[2] Stefan Grimm,[1] Paul Müller,[2] Andreas Hirsch[1]

[1] Department of Chemistry and Pharmacy, Institute of Advanced Materials and Processes (ZMP), Friedrich-Alexander-Universität Erlangen-Nürnberg (FAU), Henkestr. 42, 91054 Erlangen and Dr.-Mack Str. 81, 90762 Fürth, Germany.

[2] Department of Physics and Interdisciplinary Center for Molecular Materials Friedrich-Alexander-Universität Erlangen-Nürnberg (FAU), Erwin-Rommel-Str. 1, 91058 Erlangen, Germany.

Graphene oxide can be used as a precursor to graphene but the quality of graphene flakes is highly heterogeneous. Scanning-Raman-Microscopy (SRM) is used to characterize films of graphene derived from flakes of graphene oxide with an almost intact carbon framework (ai-GO). The defect density of these flakes is visualized in detail by analyzing the intensity and full-width at half-maximum of the most pronounced Raman peaks. In addition, we superimpose the SRM results with AFM images and correlate the spectroscopic results with the morphology. Furthermore, we use SRM technique to display the amount of defects in a film of graphene. Thus, an area of 250 x 250 µm$^2$ of graphene is probed with a step-size increment of 1 µm. We are able to visualize the position of graphene flakes, edges and the substrate. Finally, we alter parameters of measurement to analyze the quality of graphene fast and reliable. The described method can be used to probe and visualize the quality of graphene films.

Introduction

Graphene is a one atom thick nanomaterial built up by carbon atoms arranged in a honeycomb lattice. Its fascinating properties led to high interdisciplinary interest in physics, chemistry, materials science and medicine.[1-3] Especially the high charge carrier mobility, the thermal conductivity as well as the mechanical performance make new fields of applications feasible.[4] The development of applications is interlocked with the availability of the material and thus, the scalable production is a growing field of research.[1, 5-7] Despite several years of development no universal production method for graphene is found, yet. Furthermore, bulk production, structural quality and purity depend on the applied method.[1] One popular method suitable for electronic devices and transparent coatings is based on chemical vapor deposition.[8] However, the process suffers from high temperatures, the limited scalability and transfer processes, even if a very high structural quality is obtained. Other methods that are based on chemical exfoliation in liquids benefit from the cheap graphite as carbon source but suffer from small flakes size, low purity and the poor delaminating efficiency to yield monolayers of graphene.[9-11]

One method, suitable for the production of large amounts of monolayers utilizes the oxidation of graphite.[12] Initially, graphite oxide is obtained that can easily delaminate to single layers of graphene oxide (GO).[13, 14] GO is a chemically functionalized derivative of graphene and soluble in polar solvents, as water or alcohols.[15] It has attracted manifold attention in the fields of physics, chemistry and medicine because it can act as processible precursor to graphene beneath other properties.[1-3, 16] GO is highly stable as a monolayer in solution due to its chemical functionalization, namely epoxy, hydroxyl groups as well as organosulfate groups, as major species on the basal plane.[17-20]

The most popular preparation method of GO is based on the protocols introduced by Charpy and later Hummers and Offemann and relies on the oxidation of graphite in sulfuric acid with potassium permanganate as the oxidant at elevated temperature.[13, 14] Even if this method is fast and easy to manage, carbon from the honeycomb lattice is lost due to over-oxidation and thus, this conventional GO (GO-c) exhibits many permanent defects with an average distance of 1-2 nm only.[21-23] This means that the honeycomb lattice of reduced GO-c was permanently damaged. The defects limit the mobility of charge carriers and hamper the usage of the full potential of graphene in applications.[24, 25] Hence, we recently altered the synthetic protocol for the preparation of GO yielding almost intact GO (ai-GO) as a consequence of preventing the evolution of $CO_2$ during the synthesis. As a consequence, a high quality of graphene is yielded after reduction (Figure 1).[23-28] Graphene derived from ai-GO exhibits charge carrier mobility values exceeding 1000 $cm^2$/Vs for the best quality of graphene flakes and 250 $cm^2$/Vs for the mean quality of flakes.[24, 25]

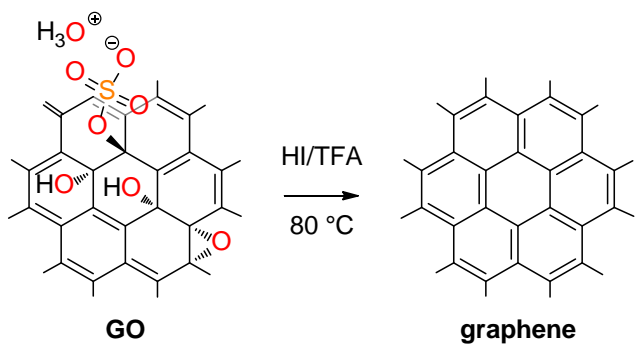

Figure 1. Reduction of ai-GO to graphene that's defect density alters between flakes.

Unfortunately, the defect density of reduced ai-GO is highly inhomogeneous and varies from flake to flake. This is a consequence of the chemical oxidation and potentially due to the heterogeneity in flake size of graphite.[9] Therefore, the defects in GO and graphene, respectively, are unevenly distributed. It turned out that Scanning-Raman-Micropscopy (SRM) is a versatile tool to determine and visualize the heterogeneity of graphene samples and to make defect densities visible, as we demonstrated for structured single flakes before.[22] Furthermore, we used SRM to investigate the efficiency of different reducing agents for ai-GO and probed the stability of the carbon framework in GO in dependence of pH and temperature.[24, 25, 29, 30] Moreover, SRM was used to determine the degree of functionalization of graphene after covalent functionalization.[31] Nevertheless, up to now no systematic SRM analysis of graphene derived from GO is available.

Here, we show how SRM can be used to properly analyze the defect density in graphene produced from ai-GO. In addition, we compare the SRM results with AFM images to correlate spectroscopy with the morphology. This study is based on the analysis of films of graphene flakes that predominantly consist of monolayers. We analyze a size of more than 60,000 µm$^2$ of graphene containing about 5,000 individual flakes by gathering Raman spectra with a step-size increment of 1 µm. At this resolution, Raman maps can be generated that are able to resolve single flakes. We identify spectral features that can be related to edges of the graphene flakes and we were able to identify few layers of graphene as well. Furthermore, we present statistical data that allow the characterization of the overall defect density of graphene films. Furthermore, we altered the increment of scanning between 1 µm and 50 µm and we reduced the scanned area to finally determine parameters that allow statistically significant analysis of graphene. This method is suitable to characterize the quality of GO after reduction and appears to be useful for comparing the defect density of differently treated graphene derived from GO.

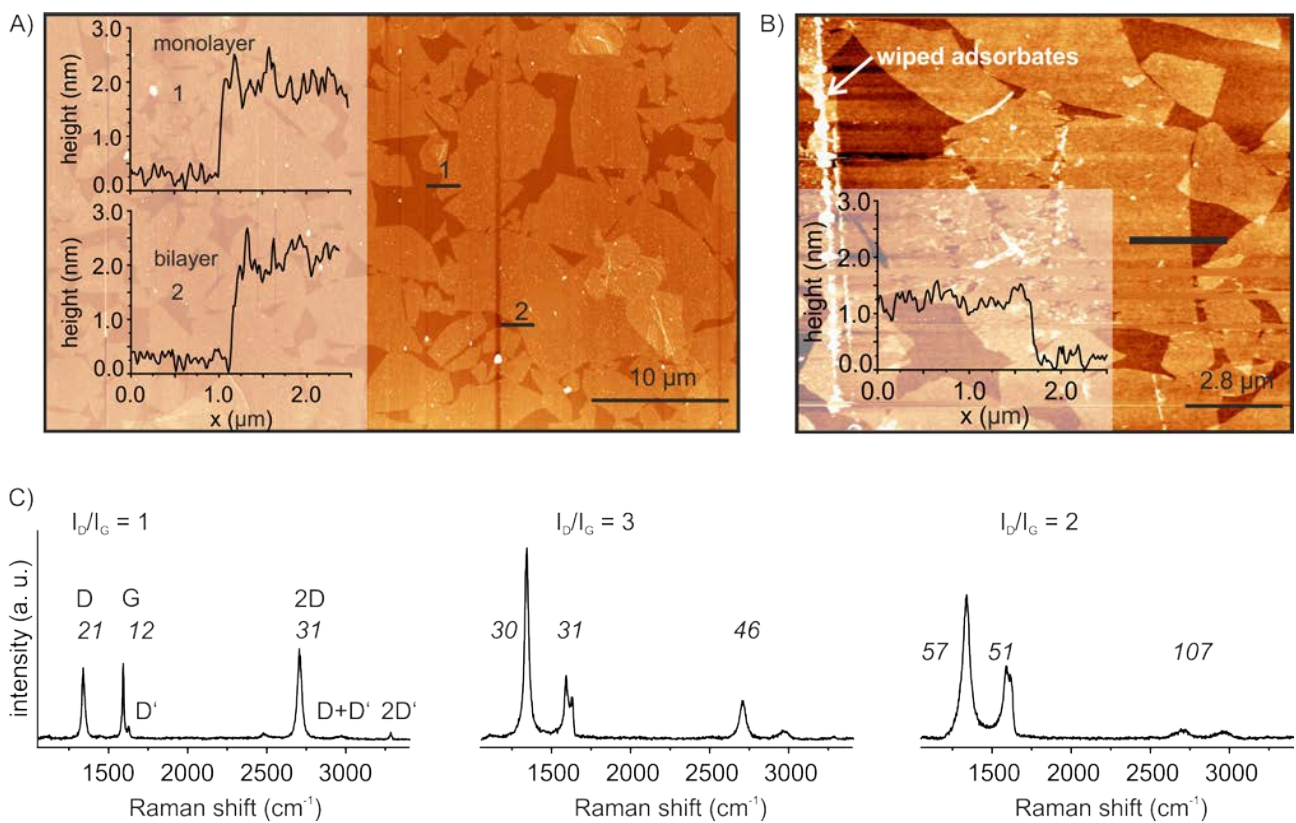

Figure 2. A) AFM image of graphene on $SiO_2$ obtained after deposition of ai-GO and chemical reduction with height-profiles of a monolayer and bilayer region; B) AFM image of a wiped area of reduced ai-GO and the height-profile; C) typical Raman spectra of reduced ai-GO with an $I_D/I_G$ ratio of 1, 3 and 2, reflecting an average distance of defects of about 11 nm, 5 nm and 1.5 nm, respectively;[32, 33] the Γ values of the D peak, G peak and 2D peak are given in italic.

The defect density in graphene can be unambiguously studied applying Raman spectroscopy.[32, 33] Raman spectra of pristine graphene display two main peaks, the G peak and the 2D peak as the most dominant one (Figure 2, Figure S3). The intensity of the G peak remains almost constant with introducing defects but the D peak evolves that is activated by structural defects. However, at a certain concentration of defects the intensity of the D peak declines and simultaneously the peak broadens (Figure 2). The correlation of Raman spectra and the distance between defects ($L_D$) was investigated by Cançado and Lucchese who introduced structural defects in graphene on the nm-scale. They studied the evolution of Raman spectra and correlated the results with the distance between defects ($L_D$) by STM.[32, 33] According to those results, the intensity of the 2D peak, second order of the D peak, declines with introducing defects and broadens continuously for this graphene. The D', D+D' and the 2D' peaks are not described in detail here because they do not contribute to the interpretation of SRM results. For this study we concentrate on the D, G and 2D peak and analyze their intensity and full-width at half-maximum ($\Gamma$). The intensity ratio of the D peak and G peak was described as a figure of merit to determine $L_D$. This $I_D/I_G$ ratio follows a relation with a maximum ratio of about 4 using 532 nm laser for excitation. The average distance of defects is about $L_D = 3$ nm for the maximum $I_D/I_G$ ratio. Therefore, the quality of graphene can be divided in graphene $L_D > 3$ nm and graphene that is dominated by defects $L_D < 3$ nm.[32, 33] Consequently, an $I_D/I_G$ ratio for example of 2 relates either to graphene or graphene that is dominated by defects.

Experimental

General Methods

Natural flake graphite was obtained from Asbury Carbon. The grade used was 3061. GO preparation from graphite was accomplished according to our previously described procedure and Langmuir-Blodgett films were prepared at 1.5 mN/m.[24] Graphene was yielded by reducing ai-GO with hydriodic acid and trifluoroacetic acid as described in the literature.[24, 25, 34] AFM imaging was performed using as Asylum Research MfP 3d AFM in tapping mode. Scanning Raman spectroscopy was accomplished on a Horiba Jobin Yvon LabRAM Aramis confocal Raman microscope with an excitation wavelength of 532 nm. The spot size was about 1 µm using an Olympus LMPlanFl 100, NA 0.80 objective in back-scattering geometry. A silicon detector array charge coupled device (CCD) was used at -70 °C for gathering Raman spectra. The spectrometer was calibrated in frequency using crystalline graphite. A motorized x,y table was used to scan the sample area using swift-mode. The sample was mounted on the x,y table and it was ensured that the focus of the laser is constant within the scanned area. The increment of scanning for the 250 x 250 µm$^2$ area was 1 µm, 0.35 s exposure time with a laser intensity of about 1 mW. For the analysis of Raman spectra, gathering and evaluation of statistical information, the preparation of histograms and creating SRM images a combination of software was used, Labspec 5, python modules from SciPy.org and OriginPro 9. The D peak and 2D peak were fitted by one Lorentz function and the G peak that is close to the D' peak was fitted by two Lorentz functions.

Results and discussion

Ai-GO was synthesized according to our recently described protocol and Langmuir-Blodgett films were prepared on the surface of Si/300 nm SiO$_2$ wafers as described before.[24] With this technique it is possible to deposit films of flakes that consist predominantly of monolayers which are an essential prerequisite for the interpretation of Raman spectra of graphene.[33, 35, 36] In this study, the reduction of ai-GO to graphene was accomplished using vapor of hydriodic acid and trifluoro acidic acid at 80 °C (Figure 1).[34] Recently, we demonstrated that this reduction method is highly efficient in terms of graphene quality and surface quality, however adsorbed species are still present on the surface.[25]

The film of graphene flakes was characterized by atomic force microscopy (AFM) and a detail of 55 x 30 µm$^2$ generated from two single AFM images is shown in Figure 2 (see also Figure S1, an AFM image of 80 x 30 µm$^2$). There are about 200 graphene flakes within an area of 50 x 50 µm² and the flake diameter varies between about 1 and 15 µm. In the inset of Figure 2A the height-profiles of monolayers and bilayers display little difference. This is due to adsorbates present after chemical reduction of ai-GO that could not be removed by common solvents.[25] However, we found that adsorbates can be removed mechanically by using AFM in contact mode.[37] With this technique adsorbates could be wiped apart, as shown in Figure 2B and Figure S2 (showing the same area before and after wiping adsorbates). Due to wiping, the height-profile reveals an about 0.7 nm reduced thickness. Assuming that impurities are also present between the substrate and graphene the estimated height is close to the theoretical thickness of graphene of about 0.4 nm. We assume that water and at least bulky sulfate that is part of the structure of ai-GO is trapped under the graphene flakes.[17]

To gain the maximum information about the defect density of graphene within the whole film we used SRM and scanned an area of 250 x 250 µm² with a step-size increment of 1 µm, about the size of the laser spot (about 1 µm²). With this increment, we analyze all flakes within the scanned area and hit many of them more than once. In addition, uncovered regions between the partially overlapping flakes are probed as well. The $I_G$ of all Raman spectra was analyzed and a histogram of $I_G$ shows the distribution of the signal intensity (Figure 3A). The maximum of the histogram can be assigned to the signal intensity of monolayers as probed by AFM (Figure 2). Furthermore, such flakes were identified as true 2D material before.[24] The continuous increase or decrease of $I_G$ apparent in Figure 2A results from measuring parts of thicker areas or uncovered areas as well. As indicated by the grey lines of Figure 3A we assign spectra with $I_{G(max)} \pm$ 20-30% to single layers of graphene while other spectra are correlated to either edges or few layers. This assignment is supported by the correlation of $I_G$ with AFM images, as depicted in Figure S4.

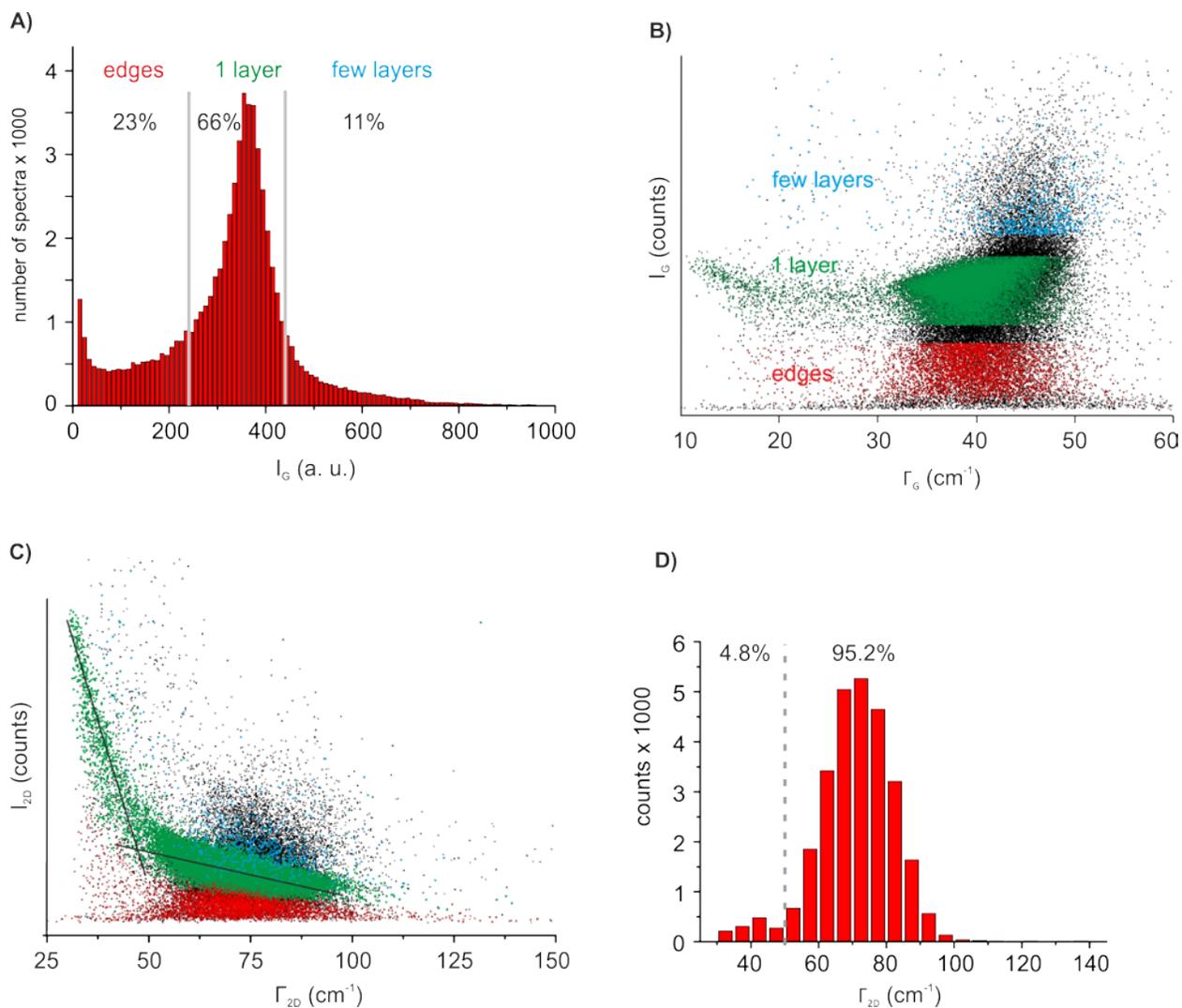

Figure 3. A) Histogram of $I_G$; B) plot of $I_G$ vs. $\Gamma_G$ extracted from more than $6 \times 10^4$ Raman spectra C) plot of $I_{2D}$ vs. $\Gamma_{2D}$ color coded: edges (red), monolayer graphene (green) and few layer graphene (blue) D) histogram of $\Gamma_{2D}$.

The plot of $I_G$ *vs.* $\Gamma_G$, as shown in Figure 3B, visualizes the heterogeneous quality of the film of graphene in terms of defect density and thickness. Since $I_G$ increases with film thickness and the distance between defects in graphene declines with the broadening of $\Gamma_G$ (from 10 cm$^{-1}$ to 60 cm$^{-1}$), it is possible to use this plot as a filter-function to identify monolayers of graphene (another criterion to assign the grey lines in Figure 3A). For very low $I_G$ values (bottom region of Figure 3B) $\Gamma_G$ between 10 and 60 cm$^{-1}$ is distributed almost equally over the whole $\Gamma_G$ region. This is due to noisy spectra recorded in the absence of graphene. Spectra with an increased $I_G$ predominantly exhibit $\Gamma_G$ values between 35 and 45 cm$^{-1}$. This region is marked in red and correlates to a combination of spectra of graphene and the substrate. As we will show below these spectra correlate to edges of graphene. The green region of Figure 3B reflects spectra that are predominantly monolayers of graphene. In this region $\Gamma_G$ = 10-50 cm$^{-1}$ and one part of these spectra with $\Gamma_G$ values between 10-30 cm$^{-1}$ reflect graphene with few defects ($L_D$ > 3 nm). Moreover, we find spectra with further increased $I_G$ values, marked in blue, that correlate to few layer graphene ($\Gamma_G$ = 40-50 cm$^{-1}$).

Since the 2D peak is dominant and sharp for monolayers of graphene only, we analyzed the $I_{2D}$ vs. $\Gamma_{2D}$ (Figure 3C) and marked the spectra of edges and few-layers that were identified by the $I_G$ vs. $\Gamma_G$ plot in red and blue, respectively. Within the green region (monolayers of graphene) an asymptotic curve progression is found that can be fitted with the following function (1).

$$I_{2D} = 79.7 + 4125^{\Gamma_{2D}} \quad (1)$$

This plot illustrates that $I_{2D}$ increases above average for $\Gamma_{2D}$ values < 50 cm$^{-1}$ as it can be extracted from the data by linear regression of $I_{2D}$ values between $\Gamma_{2D}$ = 30-50 cm$^{-1}$ and $\Gamma_{2D}$ = 70-100 cm$^{-1}$, respectively. For the quality of graphene with $\Gamma_{2D}$ < 50 cm$^{-1}$ defects do no longer interact with each other and thus, the intensity of the 2D peak increases above average.[32, 33] As a consequence the charge-carrier mobility values that we measured for such flakes before are about 250 cm$^2$/Vs.[24] This value is much higher than for graphene that is dominated by defects as indicated by a $\Gamma_{2D}$ >> 50 cm$^{-1}$.[38, 39] In Figure 3D the histogram of counted monolayer spectra according to their $\Gamma_{2D}$ is shown and we reveal that about 5% contribute to a $\Gamma_{2D}$ < 50 cm$^{-1}$ and about 95% are due to graphene that is dominated by defects.

For the evaluation of the quality of graphene the $I_D/I_G$ ratio is most often used in the literature. However, the area ratio $A_D/A_G$ is used as well.[40] Here, we compared the $I_D/I_G$ ratio and the $A_D/A_G$ using the information extracted from the 250 x 250 µm² scan and plotted the results in Figure 4. Obviously, the shape of the plot of $I_D/I_G$ vs. $\Gamma_{2D}$ and the $A_D/A_G$ vs. $\Gamma_{2D}$ are almost the same, respectively. However, the mean ratio of $A_D/A_G$ = 3.1 ±0.4 is higher than the ratio of $I_D/I_G$ = 2.7 ±0.3. Nevertheless, both plots provide the same information and are therefore both valid to characterize graphene films. We prefer to use the plot of the $I_D/I_G$ ratio because it is easier to determine intensities than peak areas.

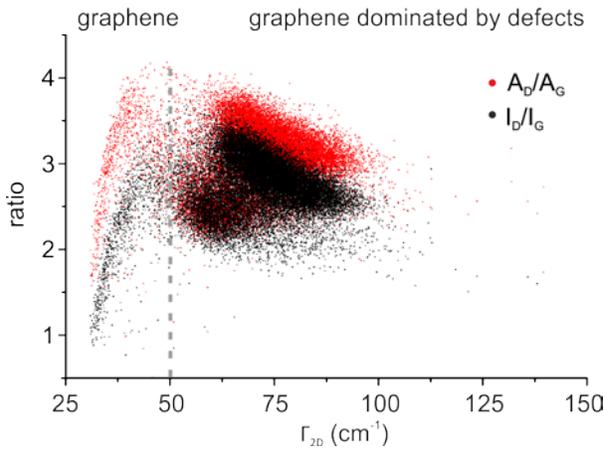

Figure 4. Plot of $A_D/A_G$ (red) *vs.* $\Gamma_{2D}$ and of $I_D/I_G$ (black) *vs.* $\Gamma_{2D}$.

In the following we interpret the spectral data to plot SRM images. In Figure 5A a color coded SRM image is shown that illustrates areas of the substrate in black, edge contribution in red, monolayers of graphene in green and few layer graphene in blue. Especially in this image the edge contribution between graphene flakes (green) and the substrate (black) becomes visible and justifies the correlation of these spectra to edge regions (red). In addition, areas that exhibit $\Gamma_{2D}$ < 45 cm⁻¹ are marked in orange that correlate to single flakes of almost intact graphene. They are almost evenly distributed over the entire area scanned by Raman spectroscopy. Thus, by plotting intensities or $\Gamma$ against the coordinates Raman spectroscopy can indeed be used as a microscopic tool.

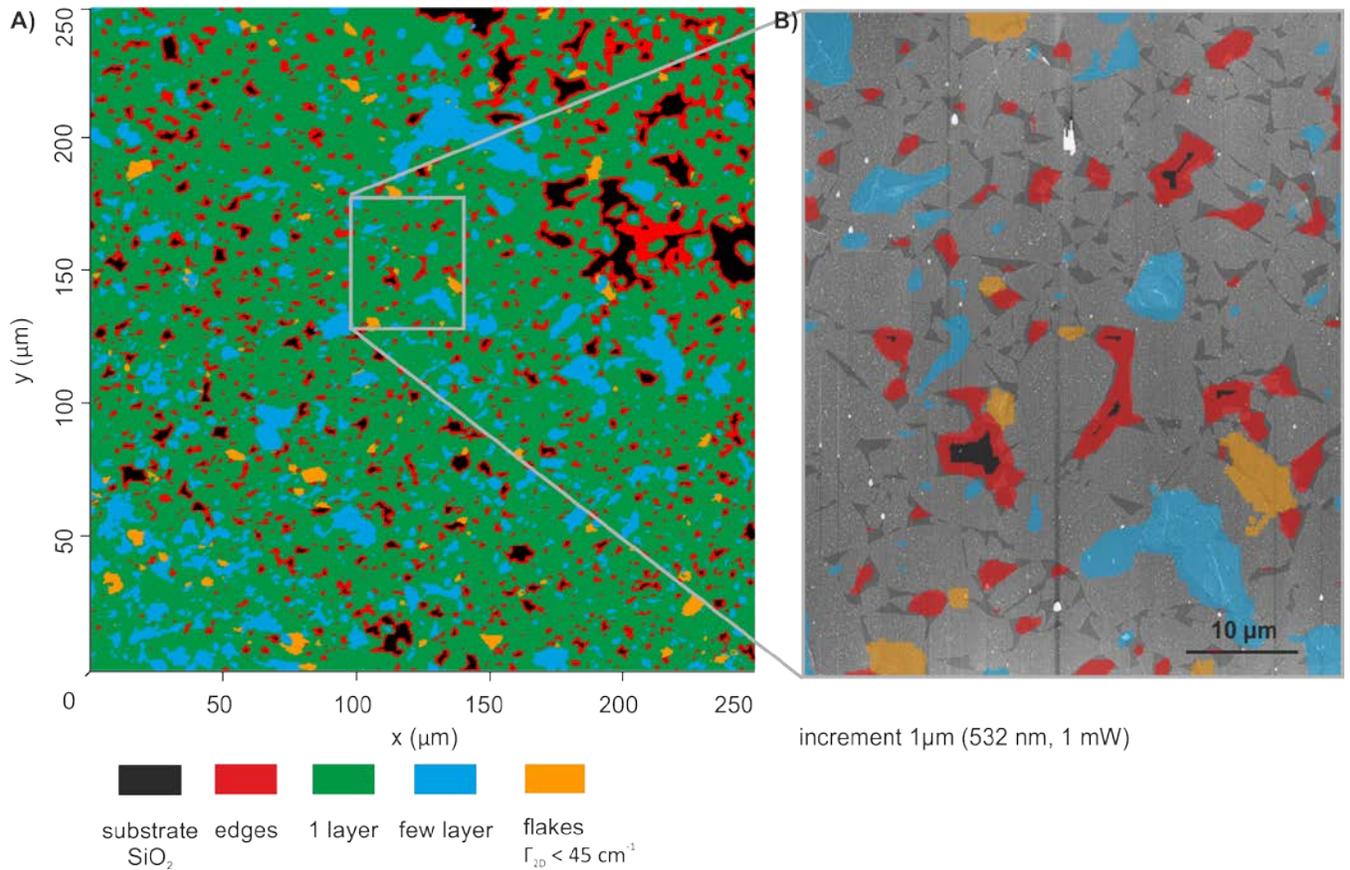

Figure 5. A) SRM image of a graphene film of 250 x 250 µm² derived from ai-GO. $I_G$ values are color-coded according to $I_G$; B) Magnification of the grey marked area of A; Raman information about the thickness of graphene flakes is illustrated as an overlay on the corresponding AFM image.

The grey marked area of Figure 5A is enlarged in Figure 5B and the SRM image is superimposed with the AFM image of the same area. The green color code for monolayers of graphene was removed to see these monolayers in the AFM image. The blue regions can be assigned to areas of few-layers of graphene while the red areas are indeed due to edges and the black areas are not covered by graphene. In addition the positions of flakes that exhibit $\Gamma_{2D} < 45$

cm$^{-1}$ are shown in orange. Due to scanning in swift-mode and a large laser-spot size the resolution is sufficient to gain information from large areas; however, the magnification reveals a limited resolution of few µm and a drift of 1-2 µm had to be corrected. Therefore, the same area was scanned again with the same increment of 1 µm but the intensity of the laser was reduced to about 0.2 mW instead of 1 mW. Furthermore, slit and hole values were optimized to minimize scattering of light and the optimized SRM image superimposed on the AFM image was prepared from these data and is shown in Figure 6.

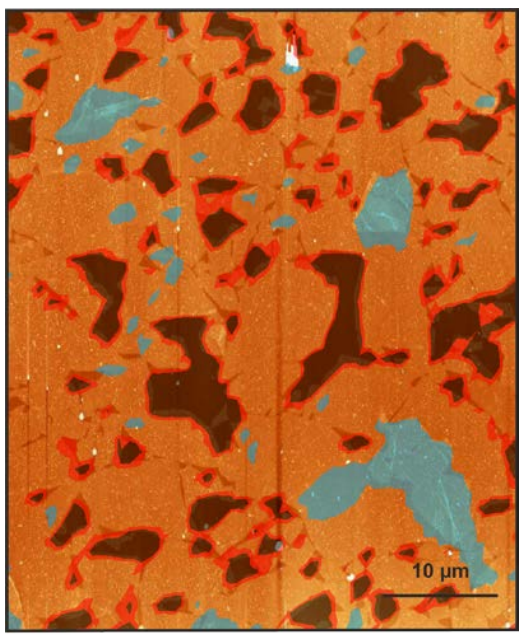

Figure 6. Raman information about the thickness of graphene flakes is illustrated as an overlay on the corresponding AFM image using 532 nm laser at 0.2 mW for Raman excitation and a step-size of 1 µm. Black: substrate; red: edges of graphene; blue: few layers of graphene.

In the next step, we evaluated the parameters that can be altered to decrease the time required for the analysis of flakes of graphene in thin films to reliable determine the quality of the sample within about 15 min. To achieve fast and reliable measurements, firstly the increment of scanning was increased stepwise from 1 µm to 2, 5, 10, 20 and 50 µm. After that, the measured spectra were analyzed and the $I_D/I_G$ values and the $\Gamma_{2D}$ were extracted. The spectra were filtered according to the $I_G$ values to exclude spectra from edges and few layer graphene, as outlined above. Histograms from the plot of $I_D/I_G$ *vs*. $\Gamma_{2D}$ were prepared and the contribution of Raman spectra with a $\Gamma_{2D} \leq 50$ cm$^{-1}$ that relate to $L_D > 3$ nm was determined. Since the contribution of graphene with $\Gamma_{2D} < 50$ cm$^{-1}$ is quite small (< 5 %), increasing the SRM increment higher than the average size of flakes results in an underestimation of this portion. Thus, an increased increment to 5, 10, 20 or 50 µm leads to an underestimated contribution of 2%, 1%, 1% or 0% of these spectra (Figure 3D, Figure 7, Figure S5). However, even if the high quality region is underestimated the average quality with a $\Gamma_{2D}$ of about 75 cm$^{-1}$ is obtained using any step-size increment between 1 and 50 µm. We conclude, for reliable analysis the increment should be smaller than 5 µm preferably between 2 and 3 µm what relates to the average size of flakes.

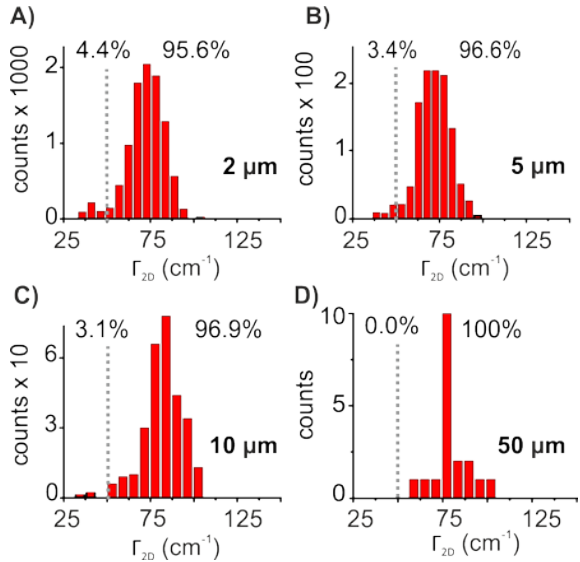

Figure 7. Histograms of $\Gamma_{2D}$ with different SRM increments A) 2 µm, B) 5 µm, C) 10 µm.

Next, the measured area was reduced to further enhance the speed of SRM, while the SRM increment was 1 µm. The original scan area of about 63,000 µm² was reduced to 17,000 µm², 8,000 µm², 4,000 µm², 1,100 µm² and 270 µm². Reliable results were obtained for the data sets with more than 1,000 spectra (Figure S6). The statistical mean values of $I_D/I_G$ and $\Gamma_{2D}$ do not significantly change and are about 2.7 ±0.3 and 72 ±10 cm$^{-1}$, respectively. The histograms of $\Gamma_{2D}$ are shown in Figure S6 and reveal that the percentage of spectra that relate to $L_D > 3$ nm are underestimated only for a scanned area < 1,000 µm$^2$. However, as displayed in Figure 5 at a certain threshold the probability increases not to measure any of the high quality flakes and thus, again the statistical analysis of the high quality region would no longer be possible.

Conclusions

We used the plot of $I_G$ *vs.* $\Gamma_G$ as a filter-function to determine $I_G$ values that can be related to edges of flakes of graphene, monolayers of graphene, few layers of graphene or the substrate. Furthermore, the plot of $I_{2D}$ *vs.* $\Gamma_{2D}$ reveals that $I_{2D}$ increases above average at $\Gamma_{2D} < 50$ cm$^{-1}$. This value, which correlates to $L_D \approx 3$ nm, was used to discriminate between graphene and graphene that is dominated by defects. We want to put emphasis on this value because not only the $I_{2D}$ value increases but also an increased charge carrier mobility values of 250 cm$^2$/Vs was found for such a flake of graphene with $\Gamma_{2D} = 50$ cm$^{-1}$.[24] Furthermore, we evaluated the difference between the $I_D/I_G$ ratio and the $A_D/A_G$ ratio statistically and confirm that both ratios can be used to describe the quality of graphene. The spectroscopic data can be used to visualize the morphology and quality of graphene and thus, microscopic images can be prepared. Here, we used exemplarily $\Gamma_{2D} < 45$ cm$^{-1}$ to show the position of graphene flakes with $L_D > 3$ nm. Furthermore, we displayed the thickness of the graphene film by using the $I_G$ value. Moreover AFM images were superimposed with SRM images to correlate the morphology with spectroscopic properties. Especially for reduced GO exhibiting an almost intact honeycomb lattice, within an average distance of defects between about 1-20 nm, the analytical method is powerful to compare the quality of differently prepared reduced GOs. Finally, we conclude that an area of 100 x 100 µm$^2$ (10,000 µm$^2$) is sufficient to gain statistically significant information about the entire film probing approximately 800 flakes of graphene within this area. The measurement of such an area is completed within about 15 min. We deduce from our results that an SRM increment smaller than the flake size should be used. Thus, depending on the film morphology we suggest to record at least 1,000 Raman spectra. In this study we get reliable results choosing an increment of 2 µm. A small area of, e.g. 50 x 50 µm may miss a significant portion of Raman spectra that relate to a high quality of graphene with $L_D > 3$ nm as it can be imagined from the SRM image shown in Figure 5. With this statistical approach chemically prepared graphene can be evaluated and compared. Therefore, new methods to graphene with almost no structural defects can be reliably identified in future studies.

ASSOCIATED CONTENT

**Supporting Information**. AFM images, Raman spectra, correlation of $I_G$ and AFM and histograms of $\Gamma_{2D}$. This material is available free of charge via the Internet at http://pubs.acs.org.

AUTHOR INFORMATION

* **Corresponding author.** E-mail address: siegfried.eigler@fau.de

Tel: +49 911 6507865005; Fax: +49 911 6507865015.

ACKNOWLEDGMENT

The authors thank the Deutsche Forschungsgemeinschaft (DFG - SFB 953, Project A1, B5 "Synthetic Carbon Allotropes"), the European Research Council (ERC; grant 246622 - GRAPHENOCHEM), and the Cluster of Excellence 'Engineering of Advanced Materials (EAM)' for financial support. The research leading to these results has received funding from the European Union Seventh Framework Programme under grant agreement n°604391 Graphene Flagship.

TOC:

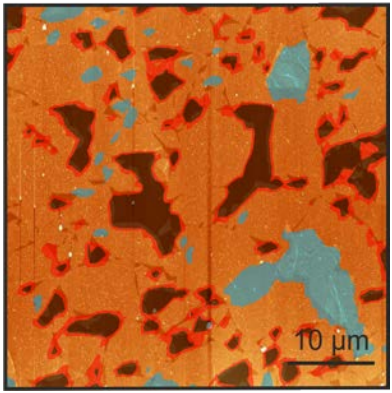

# Statistical-Raman-Microscopy and Atomic-Force-Microscopy on Heterogeneous Graphene Obtained after Reduction of Graphene Oxide

*Siegfried Eigler,*[*,1] *Ferdinand Hof,*[1] *Michael Enzelberger-Heim,*[2] *Stefan Grimm,*[1] *Paul Müller,*[2] *Andreas Hirsch*[1]

[1] Department of Chemistry and Pharmacy, Institute of Advanced Materials and Processes (ZMP), Friedrich-Alexander-Universität Erlangen-Nürnberg (FAU), Henkestr. 42, 91054 Erlangen and Dr.-Mack Str. 81, 90762 Fürth, Germany.

[2] Department of Physics and Interdisciplinary Center for Molecular Materials Friedrich-Alexander-Universität Erlangen-Nürnberg (FAU), Erwin-Rommel-Str. 1, 91058 Erlangen, Germany.

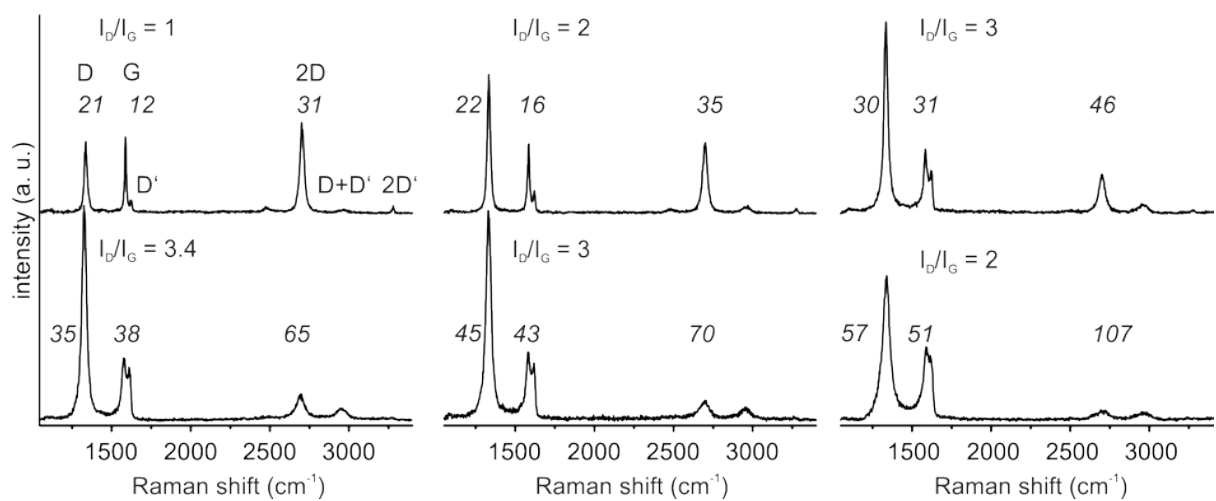

Figure S1. Raman spectra of graphene with different defect density indicated by different $I_D/I_G$ ratios obtained from ai-GO after reduction; the full-width at half-maximum ($\Gamma$) is indicated by italic numbers for the D peak, G peak and 2D peak.

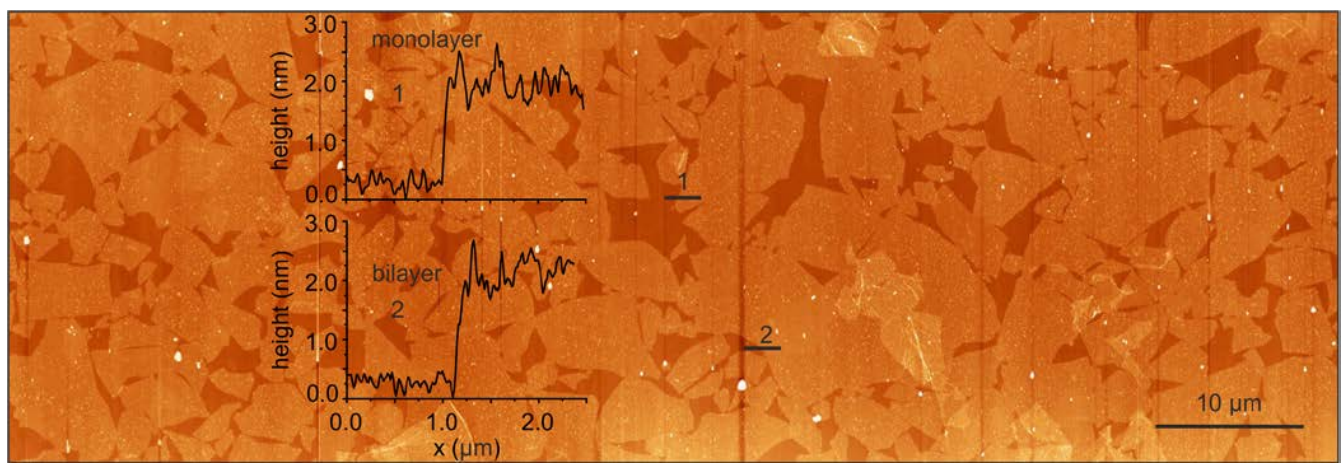

Figure S2. Stitched AFM image of a purified film of flakes of graphene obtained after reduction of ai-GO; inset: height profiles of 1: monolayer of graphene and 2: bilayer of graphene.

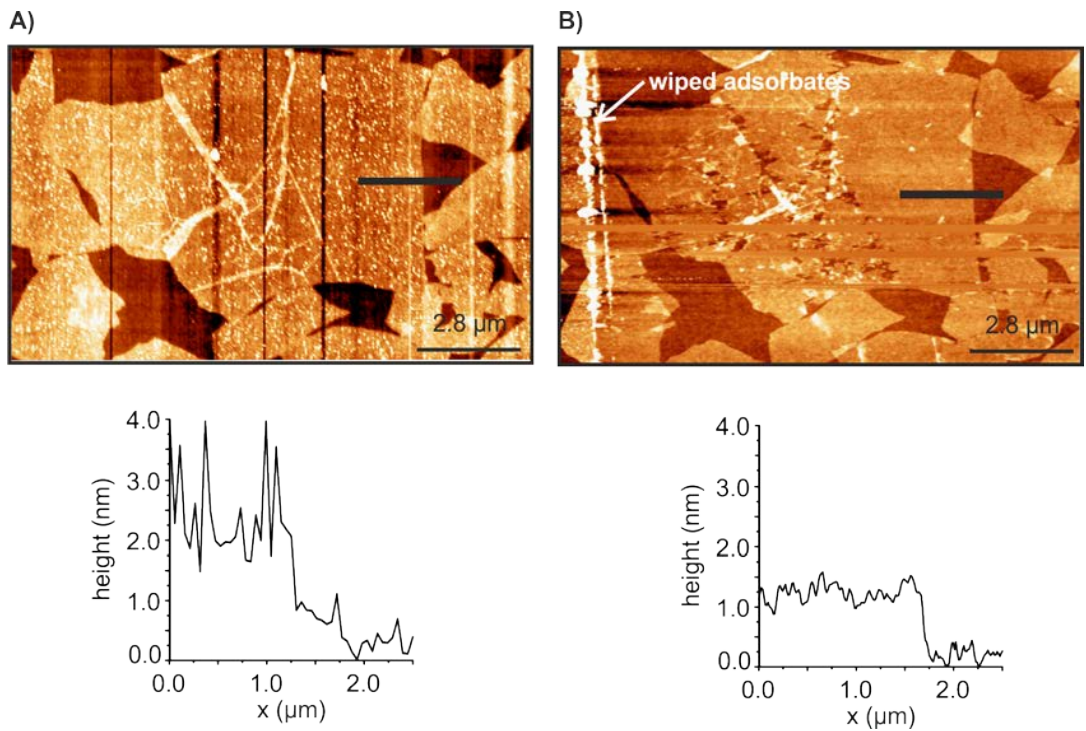

Figure S3. AFM images in tapping mode from ai-GO after reduction A) the surface of graphene is contaminated B) adsorbates have been moved apart by using the cantilever operated in contact mode; after removal of adsorbates AFM in tapping mode reveals the removal of adsorbates indicated by the reduced thickness and the thick boarder on the left of the image.

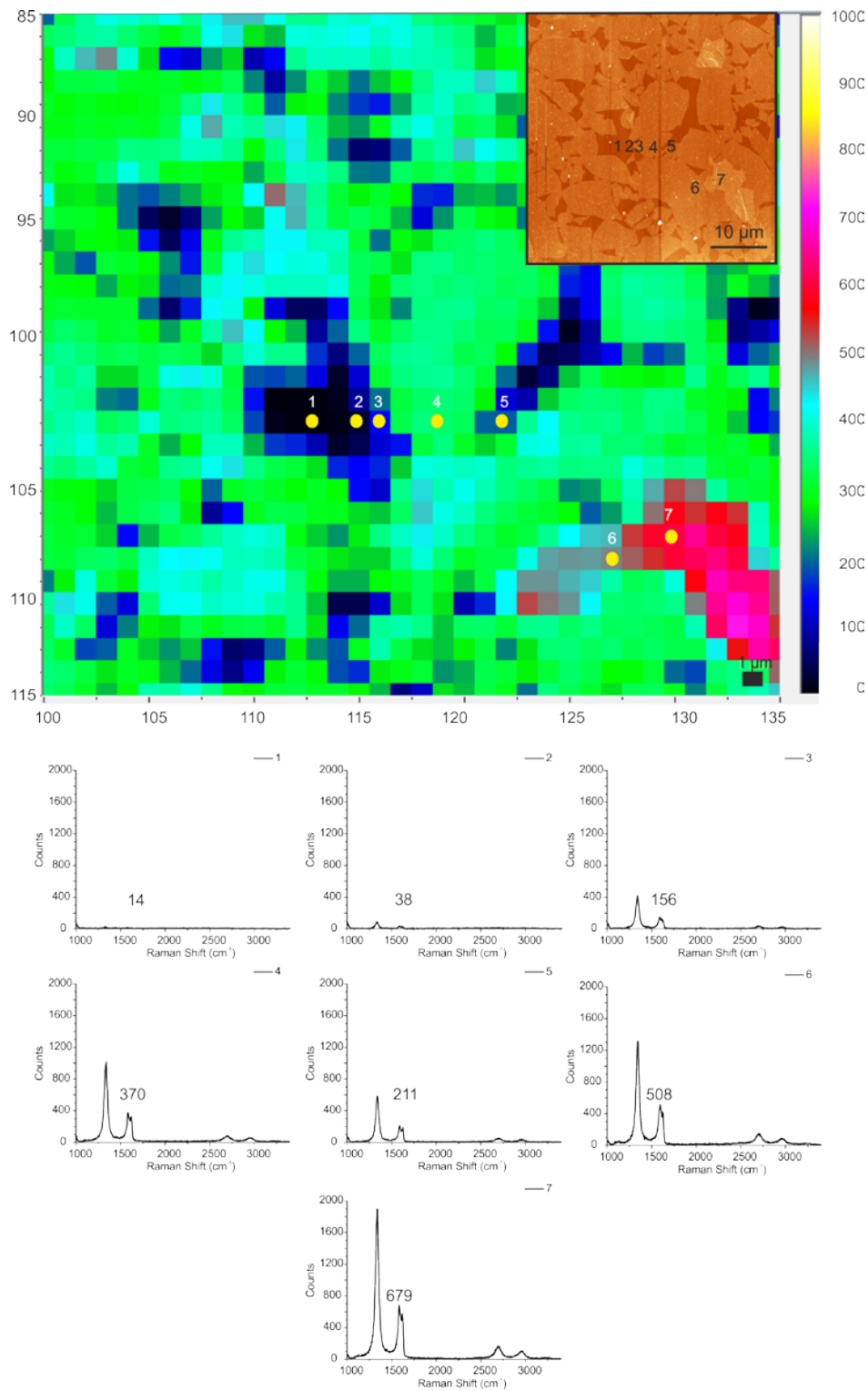

Figure S4. Mapping of $I_G$ of Raman spectra of graphene on $SiO_2$/Si. Single spectra were recorded at position 1 (substrate) 2, 3 (region of edges of flakes), 4 (single layer of graphene), 5 (region of edges of flakes), 6 (two layers of graphene), 7 (few-layer graphene); inset: AFM image with indicated position of recorded spectra; bottom: Raman spectra of position 1-7; CCD counts of the G peak are given in numbers.

S4

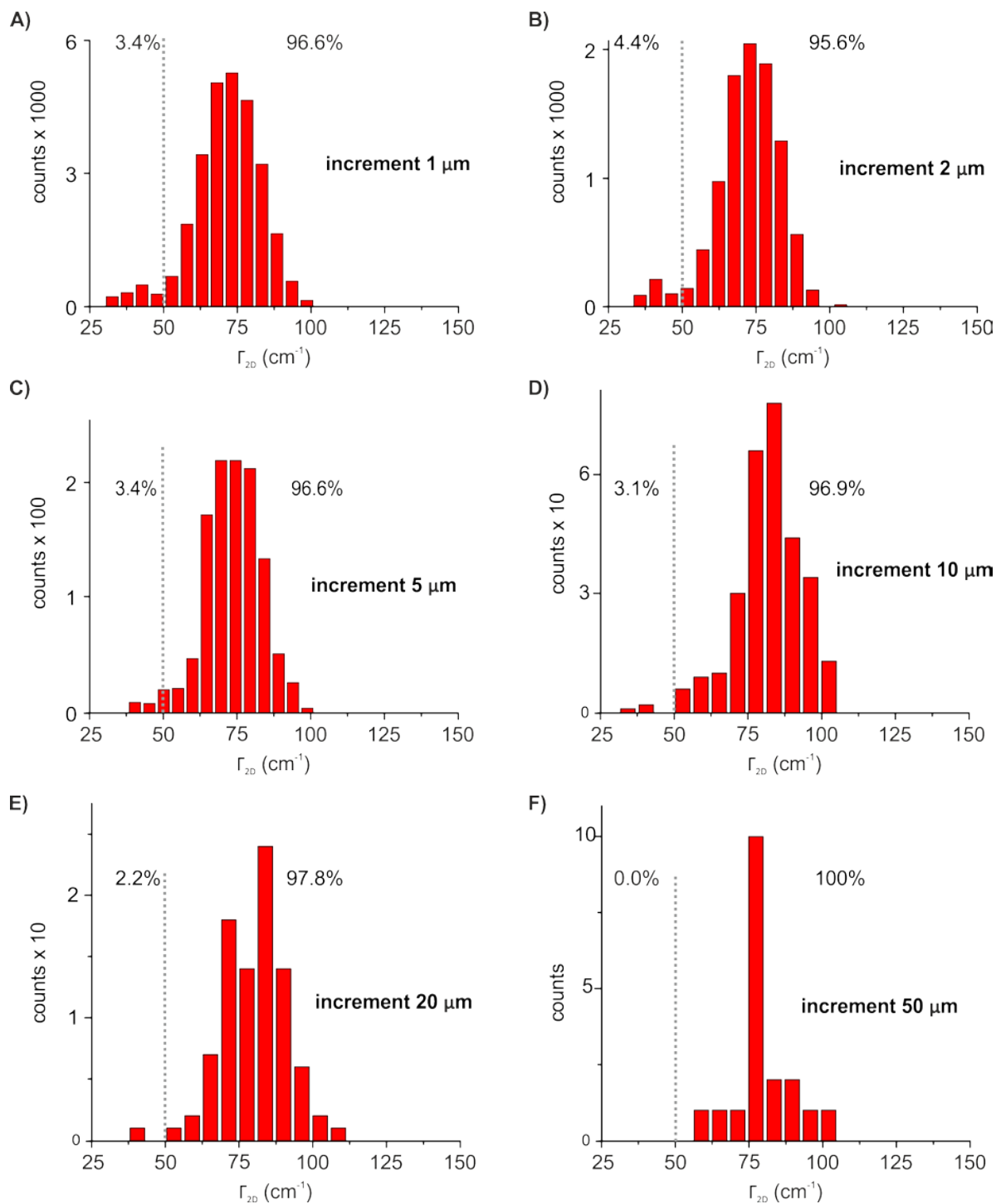

Figure S5. Histograms of $\Gamma_{2D}$ with different SRM increments A) 1 µm, B) 2 µm, C) 5 µm, D) 10 µm, E) 20 µm, F) 50 µm.

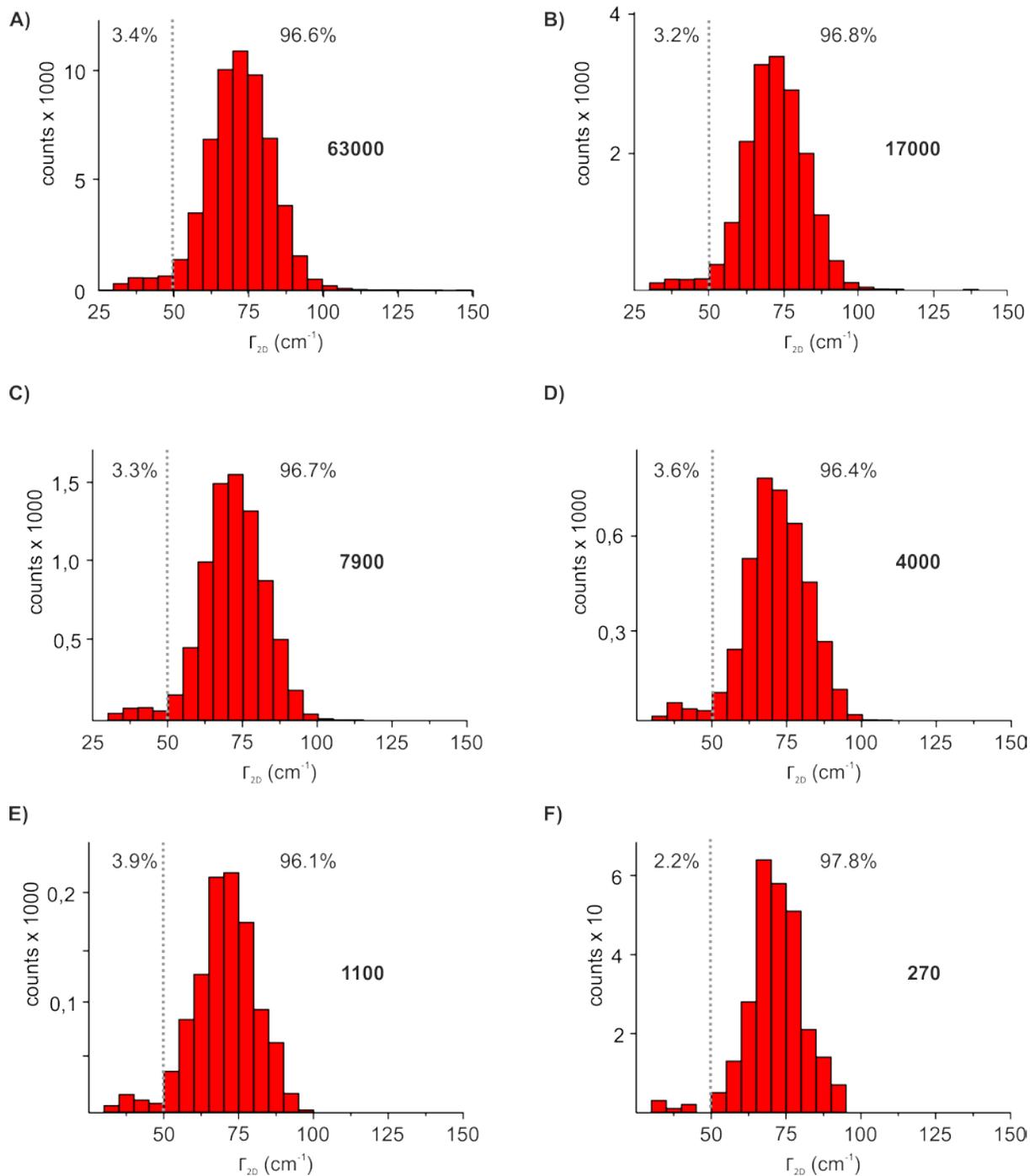

Figure S6. Histograms of $\Gamma_{2D}$ decreasing the scanned area; area was scanned by a step-size increment of 1 µm. A) 63000 µm², B) 17000 µm², C) 7900 µm², D) 4000 µm², E) 1100 µm², F) 270 µm².